\documentclass[sn-basic]{sn-jnl}% Basic Springer Nature Reference Style/Chemistry Reference Style

\usepackage{graphicx}%
\usepackage{multirow}%
\usepackage{amsmath,amssymb,amsfonts}%
\usepackage{amsthm}%
\usepackage{mathrsfs}%
\usepackage[title]{appendix}%
\usepackage{xcolor}%
\usepackage{textcomp}%
\usepackage{manyfoot}%
\usepackage{booktabs}%
\usepackage{algorithm}%
\usepackage{algorithmicx}%
\usepackage{algpseudocode}%
\usepackage{listings}%
\usepackage{pdfpages}%
\unnumbered% uncomment this for unnumbered level heads

\begin{document}

\title[Article Title]{AI-driven weather forecasts enable anticipated attribution of extreme events to human-made climate change}
%\title[Article Title]{Anticipated Attribution of Extreme Events by using Artificial Intelligence Weather Forecasts.}
%\title[Article Title]{Using Artificial Intelligence Weather Models to Accelerate Climate Change Attribution of Extreme Events}


\author*[1]{\fnm{Bernat} \sur{Jiménez-Esteve}}\email{bernatji@ucm.es}

\author[1]{\fnm{David} \sur{Barriopedro}}\email{david.barriopedro@csic.es}

\author[1]{\fnm{Juan Emmanuel} \sur{Johnson}}\email{juanjohn@ucm.es}

\author[1,2]{\fnm{Ricardo} \sur{Garcia-Herrera}}\email{rgarciah@ucm.es}

\affil*[1]{\orgdiv{Instituto de Geociencias (IGEO)}, \orgname{Consejo Superior de Investigaciones Científicas (CSIC) - Universidad Complutense de Madrid (UCM)}, \orgaddress{\city{Madrid}, \country{Spain}}}

\affil[2]{\orgdiv{Departamento de Física de la Tierra y Astrofísica}, \orgname{Universidad Complutense de Madrid}, \orgaddress{\city{Madrid}, \country{Spain}}}


\abstract{Anthropogenic climate change (ACC) is altering the frequency and intensity of extreme weather events. Attributing individual extreme events (EEs) to ACC is becoming crucial to assess the risks of climate change. Traditional attribution methods often suffer from a selection bias, are computationally demanding, and provide answers after the EE occurs. This study presents a ground-breaking hybrid attribution method by combining physics-based ACC estimates from global climate models with deep-learning weather forecasts. This hybrid approach circumvents the framing choices and accelerates the attribution process, paving the way for operational anticipated global forecast-based attribution. We apply this methodology to three distinct high-impact weather EEs. Despite some limitations in predictability, the method uncovers ACC fingerprints in the forecasted fields of EEs. Specifically, forecasts successfully anticipate that ACC exacerbated the 2018 Iberian heatwave, deepened hurricane Florence, and intensified the wind and precipitable water of the explosive cyclone Ciar\'an.
}




\maketitle

\section{Introduction}\label{sec1}

%%Key paragraph explaining the importance of attribution
Weather extreme events (EEs) such as heatwaves and tropical cyclones represent major risks to ecosystems and society \citep{ummenhofer_extreme_2017}. Anthropogenic climate change (ACC) has increased the frequency, severity and/or duration of these high-impact EEs in many areas \citep{seneviratne_weather_2021}. These changes are expected to worsen, posing potentially irreversible damage to ecosystems and new threats to humanity \citep{Kumar2023}. EE attribution raises public awareness on the impacts of ACC and provides a scientific basis to inform adaptation planning, mitigation programs and risk management for decision-making \citep{Stott2016_attribution, Otto_review_2017, NAS_2016, Knutson2017}. Enhancing our ability to predict EE and attribute them to ACC is therefore crucial to anticipate associated risks and to strengthen preparedness and resilience to future events. Despite recent developments \citep{philip_protocol_2020, Stott_2023, Faranda_Climameter_2024}, existing services struggle to answer the attribution question during the course of EE, when information is demanded. Furthermore, classical attribution only applies to specific EEs (bias selection) and how they are defined (ungeneralisable), without providing explicit attribution in the spatio-temporal pattern of the EE.

%Ability of AI models in general to predict extreme events
Artificial intelligence (AI) weather emulators offer an unprecedented opportunity to overcome these limitations by producing fast ($\sim$minutes) global forecasts at relatively high resolution on standard computers, with similar or better skill than traditional global numerical weather prediction models (NWPs) \citep{WeatherBench2}. AI-based weather models such as Pangu-Weather \citep{pangu_2023}, FourCastNet \citep{pathak2022fourcastnet}, FourCastNet-v2 \citep{bonev2023spherical} or GraphCast \citep{Lam_graphcast_2023} are competitive in simulating heatwaves, extratropical storms, or tropical cyclones \citep{pasche_validating_2024, charlton-perez_ai_2024, bouallegue_rise_2024}. By encoding information on atmospheric physical processes in their training, they simulate a physically consistent evolution of dynamical characteristics such as tropical storm formation, baroclinic waves, weather fronts, and warm conveyor belts \citep{hakim_dynamical_2024, charlton-perez_ai_2024}.

%Pseudo Global warming approach
This study builds an innovative and operationalisable approach for ultra-fast ACC attribution of EE before they occur by using a data-driven AI-based weather forecast model. As in classical attribution, the EE is simulated in a 'factual' world where it actually occurred, and in a 'counterfactual' world without ACC \citep{philip_protocol_2020, jezequel_behind_2018, CIAVARELLA_2018}. The ACC signal, as inferred from global climate models (GCMs) and the pseudo-global warming (PGW) technique \citep{brogli_pseudo-global-warming_2023}, is subtracted from the observed conditions that preceded the EE, which allows us to make AI-based reforecasts of how the EE would have been without ACC. Similar EE-constrained approaches, collectively referred to as the storyline approach \citep{shepherd_common_2016,vanGarderen_methodoloy_2021}, have been applied to attribute convective precipitation events \citep{gonzalez-aleman_anthropogenic_2023,martin_hail_2024}, tropical cyclones \citep{reed_human_influence_florence_2020,patricola_anthropogenic_2018}, extratropical cyclones \citep{ermis_event_2024} or heatwaves \citep{leach_heatwave_atribution_2021,leach_heatwave_2024}. However, they relied on NWP model simulations, which are computationally expensive, limited to the regional EE of interest and unaffordable by most users due to the need of expertise and high-performance computing. This poses challenges to achieving global ‘real-time’ attribution. 

We use an AI-based weather model, FourCastNet-v2 \citep{bonev2023spherical}, to illustrate our AI-forecast attribution approach in three high-impact EEs: a summer heatwave, a tropical cyclone, and an extratropical explosive cyclone. The results show that this approach significantly accelerates the attribution process, and lays the foundation for operational anticipatory (pre-event) attribution of EEs when applied to actual weather forecasts.

\section{Results}\label{sec2}

For each EE, we first evaluate the ability of the data-driven (AI-based) model to predict its observed evolution and characteristics (intensity, spatial pattern, etc.) by comparing it to the ERA5 reanalysis dataset \citep{hersbach_era5_2020}.  The forecasts are initialised using ERA5, which is the dataset that NVIDIA used to train the AI-based model. Counterfactual simulations are initialised from the same dates as the factual simulations but with modified initial conditions after removing the ACC signal from thermodynamic variables and adjusting geopotential height to hydrostatic balance. The ACC signal is calculated as the difference between the present (1980-2014) and preindustrial (1850-1900) climates using data from 10 historical CMIP6 simulations \citep{Eyring_cmip6_2016}. The uncertainty of factual and counterfactual forecasts of each EE is accounted for by considering an ensemble of simulations initialised at different lead times. All details related to the simulation protocol can be found in the online methods.

\subsection{Iberian heatwave (August 2018)}
An intense heatwave hit the Iberian Peninsula in 1-7 August 2018 \citep{wmo2019}. Iberia recorded the warmest week since 1950 after the 2003 heatwave, and local maxima exceeded 46$^{\circ}$C. Portugal faced the largest European fire of 2018 (27,000 ha of burnt area), and some Spanish regions reported the highest number of heat-related fatalities since 2003. The heatwave caused an increase 10\% in Iberian energy consumption and several blackouts in the suburbs of Lisbon \citep{barriopedro_exceptional_2020}.  

\begin{figure}[htb!]
\centering
\includegraphics[width=0.9\textwidth]{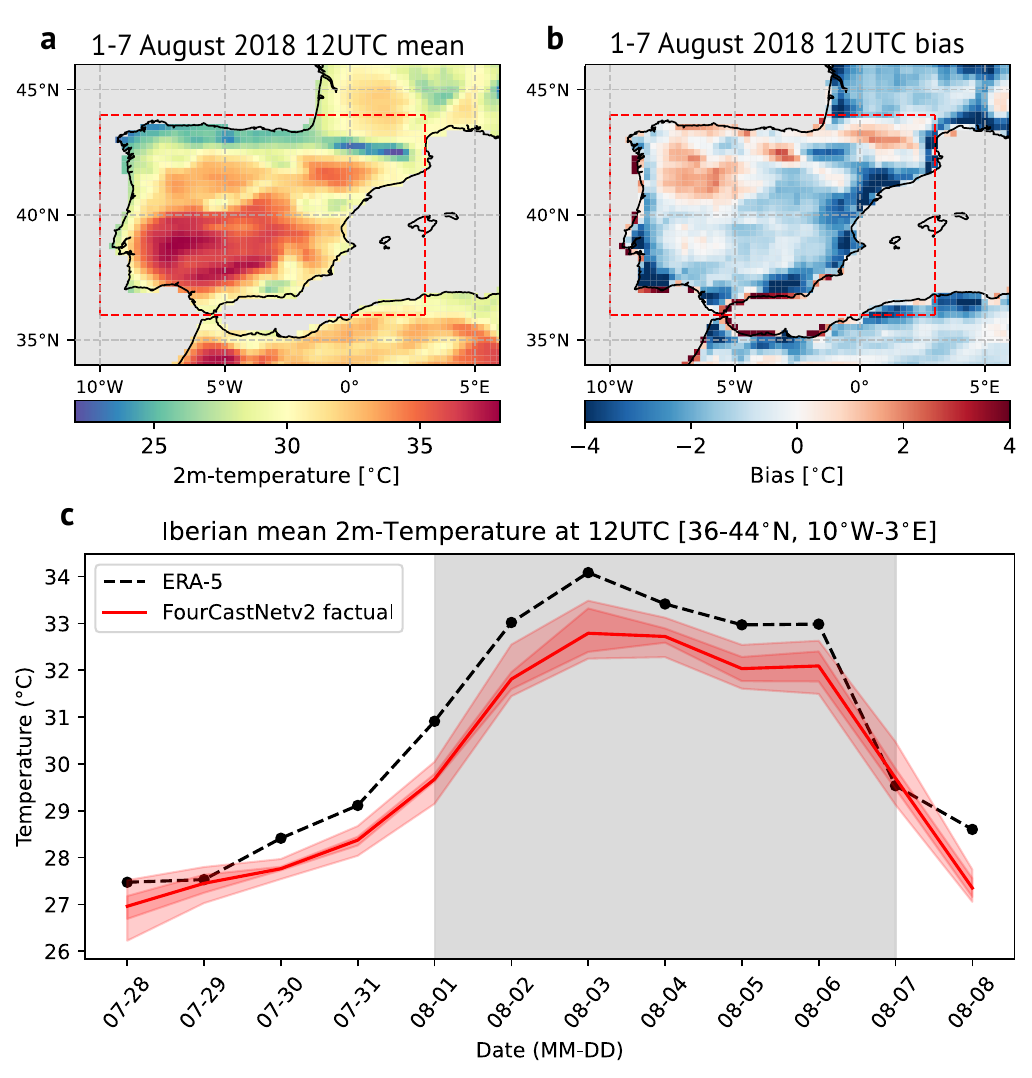}
\caption{\textbf{AI-based forecast of the Iberian heatwave of August 2018.} Geographical distribution of (a) predicted T2m and (b) bias of 12 UTC T2m with respect to ERA5 averaged over 1-7 August and 1-5 day lead times ($^{\circ}$C). (c) Evolution of the mean T2m over land Iberia ($^{\circ}$C). Red boxes in (a,b) indicate the averaging area. In (c), the dashed line corresponds to the ERA5, while the red line represents the ensemble mean. The dark red shading corresponds to the 25-75th inter-quartile range and the light red shading to the maximum-minimum range for all forecasts with lead times between 1 and 5 days. The background grey shading delimits the duration of the heatwave (1-7 August).}
\label{fig:heatwave_bias_summary}
\end{figure}

Figure \ref{fig:heatwave_bias_summary}a shows the predicted 2 m temperature field (T2m) at 12 UTC averaged during the 7-day heatwave for lead times of 1 to 5 days (the day-to-day evolution of T2m and its bias are shown in Figures S1 and S2). The model predicts the observed evolution and spatial pattern of the EE, although with a lower magnitude (averaged underestimation of $\sim$1 $^{\circ}$C), particularly in coastal areas (Figure \ref{fig:heatwave_bias_summary}b), likely due to misrepresentation of sea breeze effects. Conversely, positive T2m biases occur in northern and northwestern Spain, coinciding with the areas of greatest uncertainty (see Figure S3), which reflect the uncertain timing of the frontal system that ended the heatwave. Overall, the uncertainty associated with the date of initialisation is small (Figure \ref{fig:heatwave_bias_summary}c) and mostly increases with the lead time, as the bias does (Figure S4). The predicted geopotential height at 500 hPa (Z500) shows a small bias (Figure S5), which means that the model captures well the upper-level high-pressure system associated with the heatwave.

\begin{figure}[htb]
\centering
\includegraphics[width=1.1\textwidth]{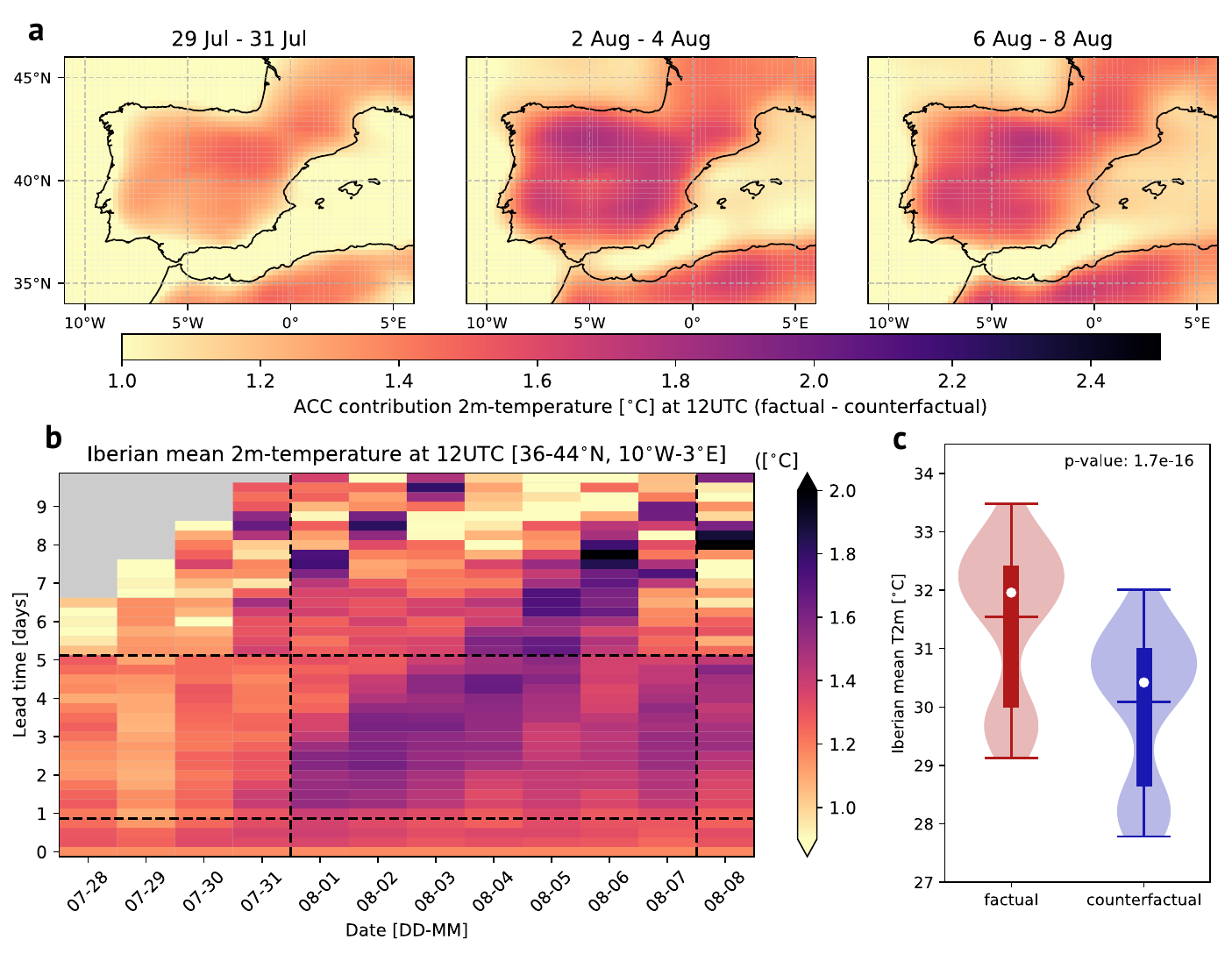}
\caption{\textbf{Climate change attribution of August 2018 Iberian heatwave.} (a) ACC fingreprint (factual minus counterfactual difference) in 12 UTC T2m averaged for 1-5 day lead times and different periods of the heatwave ($^{\circ}$C): initiation (29-31 July, left panel), peak (2-4 August, middle panel), and decay (6-8 August, right panel). (b) The area-averaged ACC signal in T2m over land areas of the Iberian peninsula ($^{\circ}$C) as a function of lead time (y-axis) and forecasted day (x-axis). (c) Factual (red) and counterfactual (blue) Violin plots (probability distribution) of the area-averaged T2m ($^{\circ}$C) for 1-5 day lead times and all days of the heatwave (1-7 August). All differences shown in (a) are statistically significant above the 95\% confidence interval. In (c), the horizontal lines represent the maximum, minimum and mean values, the white dot denotes the median of each distribution, and the vertical box the 25-75th percentile range.}
\label{fig:heatwave_pgw_summary}
\end{figure}
 
%The hybrid attribution approach allows explicit quantification of the impact of ACC on the predicted spatio-temporal pattern of the heatwave by comparing the factual and counterfactual forecasts of the EE. 
The T2m difference between factual and counterfactual forecasts tends to increase markedly within the first time step (Figure \ref{fig:heatwave_pgw_summary}b), likely reflecting the time needed by the model to accommodate the counterfactual perturbation. This ACC signal does not equal the initial perturbation, but it shows physically consistent variations in space and time (Figure S6), being stronger over land areas during the heatwave peak (Figure \ref{fig:heatwave_pgw_summary}a). 
When averaged over Iberia and the duration of the EE, these forecast-based estimates anticipate an amplification of the Iberian heatwave of $\sim$1.5 $^{\circ}$C by ACC (Figure \ref{fig:heatwave_pgw_summary}c), in good agreement with ex-post attribution statements of the EE \citep{barriopedro_exceptional_2020}.

\subsection{Hurricane Florence (September 2018)}

Hurricane Florence made landfall on the US coast as a Category 1 storm at 12UTC on 14 September 2018. Florence stalled near the coast, triggering torrential rainfall and extensive inland flooding in the southeastern US. A record-breaking rainfall accumulation of 913 mm resulting from a tropical cyclone was set in North Carolina. The storm caused \$24.23 billion in damage and 54 casualties \citep{StewartBerg_florence_2019}.

\begin{figure}[htb!]
\makebox[\textwidth][c]{
    \includegraphics[width=1.4\textwidth]{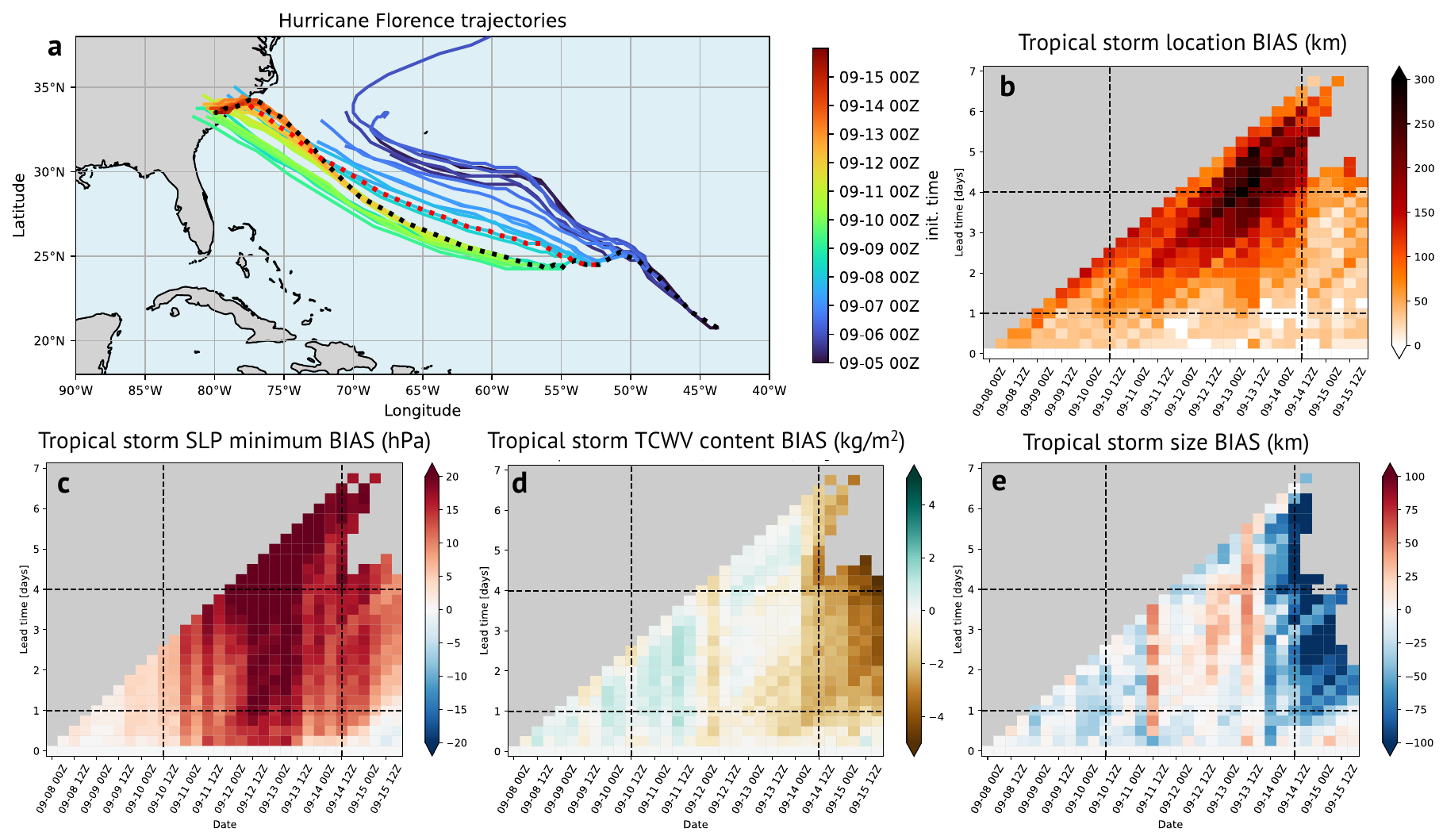}
}
\caption{\textbf{AI-based model forecast of Hurricane Florence.} (a) Forecasted trajectories of the storm's centre. Each coloured line indicates the forecasted path for the corresponding initialisation time shown on the colour bar. The observed trajectory (ERA5) is shown as a black dotted line. The red dotted line corresponds to the 00 UTC 8 September initialisation, chosen as the first initialisation date for analysis. Bias plots (model minus ERA5 difference) as a function of the lead time (y-axis) and forecasted time (x-axis) for (b) the storm's location (km), (c) intensity (minimum SLP, hPa), (d) TCWV content within a 500 km radius from the storm's centre (kg/m$^2$), and (e) storm size radius (km). See methods for the definitions.}
\label{fig:hurricane_tracks_and_bias}
\end{figure}

\begin{figure}[htb!]
\centering
\noindent\includegraphics[width=0.88\textwidth]{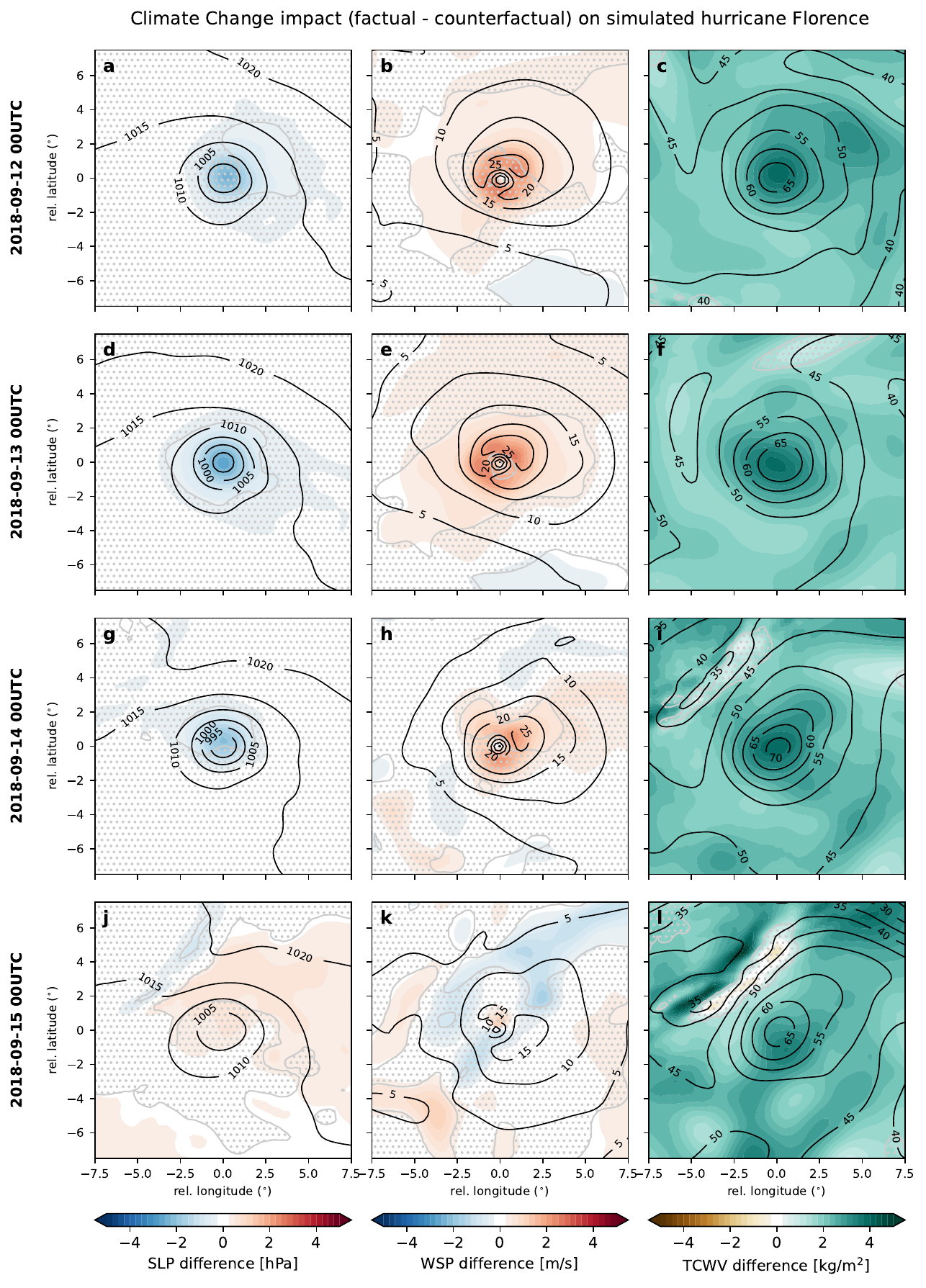}
\caption{\textbf{AI-based climate change attribution of Hurricane Florence.} Storm-centred maps of the ACC attributed change in selected variables and dates. The variables include: (a,d,g,j) SLP (hPa), (b,e,h,k) wind speed at 100 m (m/s), and (c,f,i,l) TCWV (kg/m$^2$). The dates are: (a-c) 12 September, (d-f) 13 September, (g-i) 14 September, and (j-l) 15 September, all at 00 UTC. The fields represent ensemble mean values (1-to-4 lead time averages). Contours show the total field for the lagged ensemble mean of the factual simulation, while shading shows the difference between the factual and counterfactual worlds. Non-significant values at the 95\% confidence interval according to a two-sided t-test are dotted in grey.}
\label{fig:hurricane_PGW_signal_centered_composite}
\end{figure}

%We first assess the ability of the AI-based model to predict the tropical cyclone's track by running forecasts every 6 hours from 0UTC on 5 September to 18UTC on 15 September. 
The AI-based model predicts approximately the right trajectory for initialisations after 00 UTC 8 September (Figure \ref{fig:hurricane_tracks_and_bias}a), but northward shifted tracks and too-weak cyclones for earlier initialisations. %Therefore, only forecasts initialised after 00 UTC 8 September are used.
As expected, the bias in the position of the cyclone (Figure \ref{fig:hurricane_tracks_and_bias} b)) increases with the lead time (up to 250-300 km for forecasts beyond 4 days), consistent with the typical errors of other AI-based models \citep{bouallegue_rise_2024}. The intensity, measured as the minimum sea level pressure (SLP), is underestimated by $\sim$10-20 hPa for almost all lead times, especially during the hurricane strength peak (Figure \ref{fig:hurricane_tracks_and_bias}c). This is consistent with the spatial resolution of the model being insufficient to resolve the mesoscale phenomena associated with the eye of a hurricane (see Figure 8b in \cite{bouallegue_rise_2024}). Forecasts accurately capture other characteristics such as storm size and total column water vapour (TCWV), also known as precipitable water (Figure \ref{fig:hurricane_tracks_and_bias}d,e), although they predict a smaller and drier storm after landfall (Figure \ref{fig:hurricane_tracks_and_bias}d,e). 

Despite these biases, our approach predicts ACC signals in several attributes and impact-related indicators of the hurricane, which are consistent with the thermodynamic aspects of ACC and previous attribution studies of tropical cyclones based on dynamic NWP models \citep{patricola_anthropogenic_2018,reed_human_influence_florence_2020,reed_attribution_2022}. The storm-centred composites of ACC signals in SLP, wind speed at 100 m, and TCWV are depicted in Figure \ref{fig:hurricane_PGW_signal_centered_composite}. AI-based weather forecasts anticipate a hurricane's deepening ($\sim$2-3 hPa) as a consequence of ACC (Figure \ref{fig:hurricane_PGW_signal_centered_composite}a,d,g, and Figure S7a), especially near the storm’s centre and during its peak (Figure  \ref{fig:hurricane_PGW_signal_centered_composite}d), whereas the impact of ACC diminishes after landfall (Figure \ref{fig:hurricane_PGW_signal_centered_composite}j-l). This ACC-induced intensification results in stronger wind speeds $\sim$ 2-4 m / s around the eye of the hurricane (Figure \ref{fig:hurricane_PGW_signal_centered_composite} e) and a larger storm (Figure S7c). ACC also acts to increase the amount of precipitable water up to $\sim$5 $kgm^{-2}$ surrounding the storm’s centre (Figure \ref{fig:hurricane_PGW_signal_centered_composite}c,f,l). This moistening by ACC rapidly amplifies after the initial time step (Figure S7d) and is stronger during early development than at the hurricane’s peak, suggesting a saturating effect. Conversely, the hurricane’s position was minimally affected by ACC (Figure S7b), indicating that the aforementioned ACC signals are not due to changes in the hurricane’s path.  %These findings contribute to a comprehensive and dynamically consistent depiction of a tropical cyclone’s lifecycle, intensified by ACC. Consequently, our hybrid approach can anticipate ACC signals in impact-related attributes of tropical cyclones.

\subsection{Extratropical storm Ciar\'an (October 2023)}

Extratropical Storm Ciar\'an originated over Newfoundland on 31 October 2023 and underwent explosive cyclogenesis (a pressure drop exceeding 24 hPa within 24 hours) - also known 'bomb' cyclones - as it propagated eastwards. The central pressure of the cyclone plummeted to a record low of 953 hPa, and the associated devastating winds caused widespread disruptions (flight cancellations, power outages, loss of mobile network, etc.) and at least 16 casualties across northern Europe \citep{france24_storm_ciaran_2023}. Wind gusts were accompanied by an atmospheric river - a long corridor of high precipitable water - in southern Europe, thus exemplifying a concurrent wind-precipitation EE.

\begin{figure}[htb!]
\centering
\includegraphics[width=1.0\textwidth]{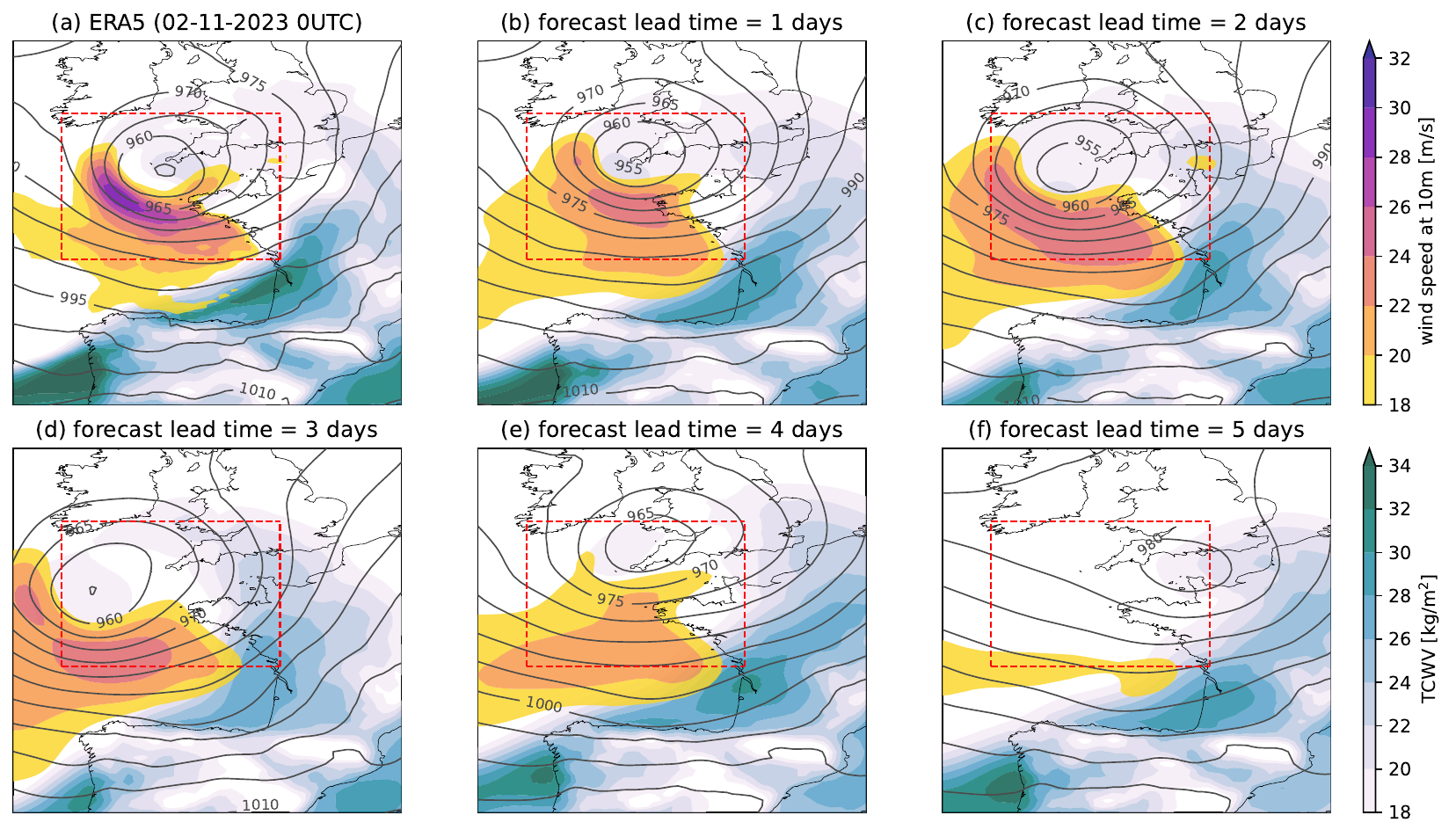} \caption{\textbf{AI-based forecast of the extratropical Storm Ciar\'an.} Maps show SLP (contours, hPa), TCWV (colour shading, kg/m$^2$) and wind speed at 10 m (overplotted colour shading, m/s) for 00 UTC 2 November 2023: (a) ERA5 reanalysis, and (b-f) AI-based forecasts for 1-to-5 day lead times.} \label{fig:extratropicalpredictability} \end{figure}

AI-based models accurately predict the wind field of the Ciar\'an cyclone \citep{charlton-perez_ai_2024}, despite a general smoothing and underestimation of wind maxima (Figure \ref{fig:extratropicalpredictability}). The atmospheric river stretching over northwestern Spain and southwestern France is successfully predicted by the model (Figure \ref{fig:extratropicalpredictability}), although with an underestimated magnitude. Short-term forecasts match the observed synoptic pattern, even predicting a deeper cyclone, but the model's performance rapidly degrades and struggles to simulate the formation of the cyclone beyond a 5-day forecast.

\begin{figure}[htb!] 
\centering
\includegraphics[width=1.0\textwidth]{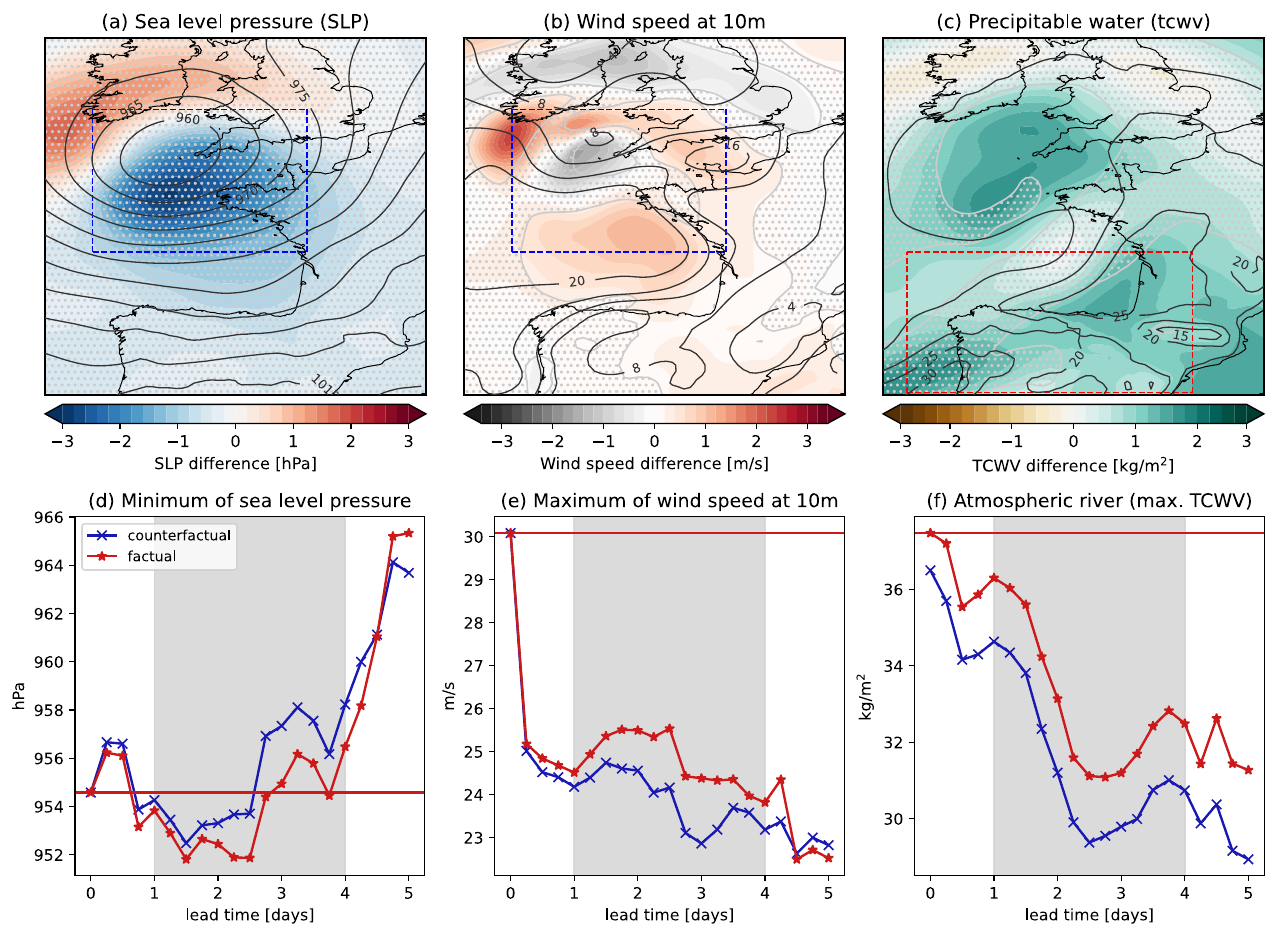}
\caption{\textbf{AI-based climate change attribution of extratropical storm Ciar\'an.} (a-c) ACC fingerprints (factual minus counterfactual difference) for 00UTC 2 November 2023 in (a) SLP (hPa), (b) wind speed at 10 m (m/s), and (c) TCWV (kg/m$^2$) fields (colour shading). In (a-c), contours show the ensemble mean (average lead times for 1-to-4 days) of the factual simulation with grey dots denoting non-significant values at the 95\% confidence interval (two-sided t-test). (d-f) The lead time dependence (x-axis) of the factual (red line) and counterfactual (blue line) forecasts of (d) local minimum SLP (hPa) inside the box of (a), (e) local maximum wind speed at 10 m (m/s) for the box in (b), and (f) local maximum TCWV (kg/m$^2$) for the box of (c). In (d-f), the horizontal red line corresponds to the value for ERA5 on 00UTC 2 November 2023, and the grey shading area is the time range used to generate the ensemble mean.}
\label{fig:extratropical_PGW_signal_centered_composite} 
\end{figure}

Our forecast-based attribution foresees an ACC-induced deepening and intensification of storm Ciar\'an (Figure \ref{fig:extratropical_PGW_signal_centered_composite}). The predicted minimum SLP is $\sim$1 hPa lower (Figure \ref{fig:extratropical_PGW_signal_centered_composite}d), and the maximum wind speeds are $\sim$1 m/s higher (Figure \ref{fig:extratropical_PGW_signal_centered_composite}e) than they would have been without ACC. An ACC-driven moistening of the storm is also predicted, more pronounced in areas prone to extreme precipitation, such as the cyclone’s centre and along the atmospheric river (Figure \ref{fig:extratropical_PGW_signal_centered_composite}c). However, the SLP signal is not statistically significant, and the same happens for TCWV in certain regions, largely due to the uncertainty in the cyclone's location and the atmospheric river (Figure S8).  

Figure \ref{fig:extratropical_PGW_signal_centered_composite}d-f shows how ACC signals vary with lead time and the variable considered. For thermodynamic variables (the maximum TCWV within the atmospheric river), it is already present at initialisation and increases up to $\sim$2 kg/m$^2$ (Figure \ref{fig:extratropical_PGW_signal_centered_composite}f). For dynamic variables (SLP and wind), the ACC signal emerges 1-2 days after initialisation (arguably denoting the time needed to accommodate the thermodynamic perturbations), and it disappears after $\sim$4 days, likely due to the predictability barrier in the storm genesis. While traditional methods have not attributed Ciar\'an to ACC, recent studies have reported similar ACC fingerprints in other extratropical cyclones \citep{ginesta_methodology_2023,ermis_event_2024}. These findings demonstrate that our approach can successfully predict ACC signals, even in dynamically-related EEs that have traditionally been challenging to attribute. %These findings highlight the potential for real-time quantification of ACC impacts on extratropical cyclones even before the impacts occur.  

\section{Discussion}\label{sec12}

%overall discussion of the advantages
This article presents a hybrid approach to attribute EEs to ACC by comparing AI-based weather forecasts from the factual world, where the EE unfolds, with those from a counterfactual world, in which human influences are altered in a physically consistent way. This methodology leverages a blended climate-weather perspective and provides several advantages over similar approaches applied to NWP models. Firstly, it delivers global predictions and ex-ante attribution of various types of high-impact weather EEs simultaneously. Being ultrafast and accessible to all, this tool copes with the long-standing demand for timely attribution of EEs \citep{Stott_2023} since the attribution is done in forecasted rather than observed outcomes. By providing global 6-hourly fields of multiple variables in the two worlds, the method yields explicit attribution in the evolving spatio-temporal pattern and impact-related variables of the EE. This avoids some ‘framing’ choices of traditional methods, such as the definition of EE, which can affect the attribution result \citep{NAS_2016}. Importantly, this approach enables global attribution, facilitating the attribution of simultaneous EEs across regions.  

%dicsussion of the prediction and biases of the model
The versatility and generalisability of the method were demonstrated in three case studies that encompass different types of high-impact EEs and impacted regions. They include challenging EEs to attribute because of their small spatio-temporal scales and/or low signal-to-noise ratios. The AI-based model forecasts successfully predicted the three EEs, sometimes more than 4-5 days in advance, albeit with a general underestimation of the magnitude. This is a common bias of AI-based models and state-of-the-art physics-based models (e.g. the IFS; \cite{pasche_validating_2024}), likely due to insufficient resolution to resolve small-scale processes and their interactions \citep{Zappa_cmip5_extratropical_2013, li_extratropical_2014, patricola_anthropogenic_2018}. While some studies have explored this issue (e.g., \cite{bellprat_towards_2019}), further research is needed to fully understand how model biases affect the quantification of ACC signals.  
%These predictions aligned well with ERA5 reanalysis data, particularly for short lead times (1-4 days).

%summary of the ACC impact on EE 
%Our methodology illustrates three examples of successful detection of ACC signals in the forecasted, rather than the observed, fields of EEs. 
Certain aspects of these EEs have already been attributed to ACC with classical methods, including dynamical models. Our approach yields consistent physically grounded attribution statements before the EE, and broadens the attribution to 2D instantaneous multivariate fields on global scales. We find that ACC made the 2018 Iberian heatwave $\sim$1.5 $^{\circ}$C warmer than if it had occurred in the preindustrial period. This ACC signal varies in space and time, suggesting that the model has learnt to reproduce amplifying effects such as land-atmosphere feedbacks \citep{Miralles2019, barriopedro_heat_2023,domeisen_prediction_2023} without explicit simulation. Hurricane Florence is also predicted to be stronger, moister, and slightly larger than it would have been in a preindustrial climate, consistent with an intensification of tropical cyclones under global warming \citep{reed_human_influence_florence_2020,patricola_anthropogenic_2018,knutson_tropical_2020}. ACC also contributed to the pressure deepening and wind intensification of the Ciar\'an storm. The impact of ACC on extratropical cyclones is still debated due to counteracting effects \citep{catto_future_extratropical_2019,li_extratropical_2014, seiler_how_2016, priestley_future_2022}, this lack of understanding being reflected in a low number of attribution studies. In spite of this, there is consensus that ACC will increase the precipitation associated with cyclones \citep{kodama_new_2019, catto_future_extratropical_2019}, consistent with the ACC-induced moistening of the Ciar\'an storm. 

%In this sense it is worth of noticing that there is still a marked geographical bias in the EEs that have been attributed to ACC. The fast rise in the frequency and intensity of many types of EEs also increases the chances of concurrent EEs over different regions of the world (e.g. the 2018 heatwave), questioning the viability (and adequacy) of attributing single events in isolation. 

Our methodology has proven effective in delivering anticipated attribution of weather EEs, but there are opportunities for improvements. An ideal approach would use an ensemble of probabilistic predictions (e.g. stochastic physics, \cite{berner_stocastic_2017}) and of multiple AI-based models, quantifying different sources of uncertainty \citep{Lavin-Gullon_multiphysics_2021}. New AI-based forecast ensembles are in development \citep{price_gencast_2023,jeppesen_aifsens_2024}, although not yet open source. Meanwhile, the lag-based ensemble approach \citep{brenowitz_practical_2024} used here provides a simple approach to account for the uncertainty in the predictability of EEs associated with the initial conditions. Furthermore, we conducted the same attribution exercises using AI-based weather forecasts from Pangu-Weather \citep{pangu_2023}, obtaining comparable results (see Figures S9 to S11). However, some discrepancies emerge, which may stem from differences in the model variables (e.g., Pangu-Weather uses specific humidity instead of relative humidity, and omittes TCWV) and architectures. Another limitation is that forecast-based attribution is restricted to weather-related EEs and associated impacts (wildfires, etc.), but cannot attribute climate EEs with time scales beyond the meteorological predictability (e.g., agricultural droughts). Being constrained by observed atmospheric states alone, our approach yields partial attribution of ACC influences on EEs since it does not explicitly account for ACC-induced changes in slowly-varying components like sea surface temperatures (SSTs). Including the effects of altered boundary conditions (e.g. SST and sea ice cover) on the attribution of EEs could be tested in hybrid climate models that use AI to improve and speed up parameterisations with physics-based numerical methods for resolved processes \citep{kochkov_neural_2024}.

Effective attribution would deliver timely assessments of EEs and regular updates upon demand, avoiding ad hoc selection and uneven geographical coverage of EEs \citep{Stott_2023, NAS_2016}.  Our approach represents an unprecedented opportunity for operational early attribution of EEs. While forecast-based attribution is also possible with NWP models, operational implementation faces challenges. Dynamical simulations are regional (EE-oriented), computationally expensive, and time and personnel-demanding. Therefore, they are only affordable for a few research teams / centres (certainly not those from vulnerable underdeveloped countries) and cannot address concurrent EEs, which affect distant regions simultaneously, posing threats to multiple sectors worldwide \citep{Kornhuber2020, jimenez-esteve_heat_2022}. Although forecasts are initialised here with ERA5 reanalysis, which is available with a five-day delay, real-time weather forecasts initialised with the Integrated Forecast System (IFS) analysis enable actual attribution of EEs before they occur. Indeed, there are no apparent differences in the attribution of instantaneous global fields between ERA5 and IFS forecasts (Figure S12), showcasing that operational anticipatory global attribution of EEs is now achievable. This pioneering ultrafast AI-based method also has the potential to transform our capabilities to attribute the impacts of EEs before they occur, providing a valuable tool for adaptive planning in response to the growing threat of EEs.

\backmatter

\section{Methods}\label{sec:methods}

\subsection{AI-based model and EE forecasts}

We use FourCastNet-v2 \citep{bonev2023spherical}, an improved version of FourCastNet~\citep{pathak2022fourcastnet} developed by Nvidia and collaborators. This global data-driven weather forecasting model provides short to medium-range global predictions at 0.25$^{\circ}$ horizontal resolution for 13 pressure levels (1000, 925, 850, 700, 600, 500, 400, 300, 250, 200, 150, 100, 50 hPa) at 6-hour intervals. For training and initialization, FourCastNet-v2 uses five variables at different pressure levels [temperature (t), zonal wind (u), meridional wind (v), geopotential height(z), relative humidity (r)] and eight surface level variables [2m-temperature (t2m), surface pressure (sp), sea level pressure (slp), wind components at 10 meters (u10,v10) and 100 meters (u100,v100), total column water vapor (tcwv)], which are also forecasted. This AI-based model matches the forecasting accuracy of the ECMWF Integrated Forecasting System (IFS), a state-of-the-art Numerical Weather Prediction (NWP) model, for large-scale variables in the short-term, and even outperforms IFS for variables with complex fine-scale structure \citep{bonev2023spherical}.

FourCastNet-v2 builds on its predecessor FourCastNet by incorporating spherical Fourier Neural Operators (FNO), which improve forecast accuracy and long-term stability by enforcing physical consistencies with the spherical geometry of the Earth. This emulator is trained in two stages: a single autoregressive step (6-hour predictions) followed by fine-tuning with multiple autoregressive steps. The loss function is a simple weighted mean squared error over the global field. Although the entire training procedure is prohibitively expensive for most users (four hours with eight NVIDIA DGX machines), the weights are freely available for FAIR use \citep{earth2mip}, allowing users to fine-tune the model for different grid resolutions. In addition, inference is very fast, for example, it takes 13 minutes to generate a global forecast for one year on a single NVIDIA RTX A600 GPU.

The model was trained using the ERA5 reanalysis \citep{hersbach_era5_2020} from 1979 to 2015, with 2016 and 2017 used for validation, and 2018 for testing. Thus, we select extreme events (EE) that occurred from 2018 onwards, outside the training and validation period. For each EE, we produce 15-day forecasts for various lead times, from several days before the EE (depending on its predictability) until the end of the EE. In these factual simulations, the model is initialised from the ERA5 reanalysis dataset \citep{hersbach_era5_2020}. Then, a lagged ensemble of predictions is generated with lead times between 1 and 4-5 days, similar to the method described in \cite{brenowitz_practical_2024}. This ensemble of factual simulations is used to compute an unweighted ensemble mean and to assess the uncertainty associated with the model initialisation time. 

%For our application, we intend to use it to do multiple runs over a 10 day period.

%In this study, we use FourCastNet-v2 \cite{bonev2023spherical} is a global data-driven weather forecasting model developed by Nvidia and collaborators that provides accurate short to medium-range global predictions at 0.25$^{\circ}$ horizontal resolution. FourCastNet-v2 builds on his predecessor FourCastNet \cite{pathak2022fourcastnet}, short for Fourier Forecasting Neural Network, by implementing spherical Fourier Neural Operators (FNO) to capture spatial dependencies better, enhancing the precision of its forecasts. FourCastNet accurately forecasts high-resolution, fast-timescale variables such as surface wind speed, precipitation, and atmospheric water vapor. FourCastNet-v2 matches the forecasting accuracy of the ECMWF Integrated Forecasting System (IFS), a state-of-the-art Numerical Weather Prediction (NWP) model, at short lead times for large-scale variables, while outperforming IFS for variables with complex fine-scale structure. FourCastNet-v2 generates a week-long forecast in less than a few seconds, orders of magnitude faster than IFS. he model was trained on a subset of ERA5 reanalysis data on single levels and pressure levels.

\subsection{Climate Change Attribution Method.}

The method removes the ACC signal from the initial conditions of the factual simulations using the PGW approach \citep{brogli_pseudo-global-warming_2023}. Firstly, monthly averages for various thermodynamic variables are derived from historical (1850-2014) simulations of 10 CMIP6 (Coupled Model Intercomparison Project Phase 6, \cite{Eyring_cmip6_2016}) GCMs (see Table \ref{table1}). For each GCM, only one realisation is used. These considered variables include 2 m temperature (2D, t2m), total column water vapour (2D, tcwv), air relative humidity (3D, r), and air temperature (3D, t). To maintain hydrostatic balance, the geopotential height is adjusted in accordance with the modified variables. It is important to ensure that pressure level variables have no missing values (e.g. under topography) since the AI-based model requires complete datasets. All CMIP6 data are interpolated into a common regular lat/lon 2.5x2.5$^{\circ}$ grid.

For each thermodynamic variable $X$, with $X=(t2m,tcwv,t,r)$, the ACC signal ($\Delta X_{ACC}$) is defined as the difference between the factual (1980-2014) and counterfactual (1850-1900) periods:
\begin{equation}
\Delta X_{ACC} (m,t,x,y,p) =  \overline{X_{1980-2014}} (m,t,x,y,p) -   \overline{X_{1850-1990}} (m,t,x,y,p)    
\end{equation}
 
Here, the overbar denotes the multi-year mean, $m$ the CMIP6 model, $t$ the calendar month of the climatology, ($x$,$y$) the grid point, and $p$ the pressure level. The multi-model mean (MM) ACC signal is then computed for each thermodynamic variable $X$. Finally, the monthly series $\Delta X_{ACC,MM} (t,x,y,p)$ is linearly interpolated to the exact initialization date of the counterfactual simulations (by taking the monthly means as representative of the 15th day of each month). 

The initial conditions of the counterfactual simulations are obtained by subtracting the thermodynamic perturbations from the initial conditions of the factual runs ($X_{PGW}=X_{ERA5} - \Delta X_{ACC,MM}$), and adjusting the geopotential height to be in hydrostatic balance with the introduced perturbations. The geopotential height perturbation ($\Delta\phi_{ACC}$) is calculated as the difference between the hydrostatically adjusted geopotential of the factual conditions (ERA5) and counterfactual ($X_{PGW}$) conditions. To compute the hydrostatic geopotential, we integrate pressure levels, from bottom to top, using the following equation (based on equation 6 in \cite{brogli_pseudo-global-warming_2023}):
\begin{equation}
\phi_{k+1} = \phi_k + (\ln(p_k) + \ln(p_{k+1})) R \frac{(T_{v,k} + T_{v,k+1})}{2}
\end{equation}
for $k=1,...,13$. Here, $k$ is the pressure level (where $k=1$ corresponds to $p_1 = 1000 hPa$), $R = 287 J kg^{-1} K^{-1}$ the ideal gas constant, and $T_{v,k}$ the virtual temperature at pressure level $k$, computed using $T_k$ and $r_k$. The $\Delta \phi_{ACC}$ is then subtracted from the ERA5 initial conditions before running the counterfactual simulations. 
 
Once the initial counterfactual conditions are obtained, we repeat the same procedure as used for the factual simulations. This includes generating a lag-based ensemble of counterfactual forecasts for the same initialisation dates and lead times as the factual runs. The ACC signal is then defined as the difference between the lagged ensemble mean of the factual and counterfactual simulations.

\begin{table}[htb!]
\caption{CMIP6 models used for calculating the climate change deltas of thermodynamic variables.}
\centering
\renewcommand{\arraystretch}{1.2} % Add extra vertical spacing between rows
\begin{tabular}{p{3cm}p{5.5cm}p{4cm}}
\hline
  \multicolumn{1}{c}{Model acronym}  & \multicolumn{1}{c}{Name} & \multicolumn{1}{c}{Reference} \\
 \hline
   ec-earth3-veg-lr & EC-Earth3 - Vegetation - Low Resolution &\cite{EC-Earth3-CMIP6-2022} \\
   ec-earth3-cc  & EC-Earth3 - Carbon Cycle &\cite{EC-Earth3-CMIP6-2022} \\
   ec-earth3-veg & EC-Earth3 - Vegetation &\cite{EC-Earth3-CMIP6-2022} \\
   awi-cm-1-1-mr & Alfred Wegener Institute - Climate Model - Medium Resolution & \cite{AWI_CMIP6_2020}   \\
   canesm5-1  & Canadian Earth System Model version 5 &  \cite{CanESM_CIMP}  \\
   cmcc-cm2-sr5  & CMCC - Climate Model 2 - System Resolution 5 & \cite{CMCC-ESM2-CMIP} \\
   cmcc-cm2-hr4  & CMCC - Climate Model 2 - High Resolution 4 & \cite{CMCC-CM2-HR4} \\
   cmcc-esm2 & CMCC - Earth System Model 2 & \cite{CMCC-ESM2-CMIP}  \\
   bcc-csm2-mr  & Beijing Climate Center - Climate System Model - Medium Resolution & \cite{BCC-CSM2MR}  \\
   cams-csm1-0 & Chinese Academy of Meteorological Sciences Climate System Model 1.0 & \cite{CAMS-CSM1.0} \\
 \hline
 \end{tabular}
 \label{table1}
 \end{table}

\subsection{Tropical cyclone tracking and metrics}

To analyse tropical cyclone Florence, we identify and track the storm centre as the local minimum of sea level pressure (SLP) in the region [18-40$^{\circ}$N,85-40$^{\circ}$W]. For ERA5 and each initialisation, the hurricane's position is determined every 6 hours using the {\textit{minimum\_filter}} function of \textit{scipy} python package \citep{2020SciPy-NMeth} with a neighbourhood size of 100 points. To maintain a consistent trajectory, the distance between consecutive tracked points should be the closest to all detected SLP local minima using \textit{minimum\_filter} function and must be displaced less than 250 km within each 6-hour time step. We also record the SLP minimum used to define the storm's centre.

To calculate storm-centred composites (Figure \ref{fig:hurricane_PGW_signal_centered_composite}), we extract a 61x61 grid region (15x15$^{\circ}$) around the storm's centre for SLP, wind speed at 100 m and TCWV. Additionally, we compute the Euclidean distance from each point to the storm's centre in order to average the wind speed as a function of the storm radius. The storm size is then defined as the distance from the storm's centre to the radius where the average wind speed is 10 m/s as in \cite{reed_human_influence_florence_2020}. We also define a water vapour content index as the TCWV averaged within a radius of 500 km of the storm's centre. 

\subsection{Statistical significance of the results}

To assess the statistical significance of the ACC attribution signal in the 2D fields of the various variables analysed, we employ a two-sided t-test using the \textit{ttest\_ind} function from the {\textit{scipy}} Python package \citep{2020SciPy-NMeth}. This test assumes that the two distributions of the factual and counterfactual worlds have equal variances and are independent of each other. For each distribution, the population contains all AI-based forecasts initialised between 1 and 4 days (5 days for the heatwave case) before the EE.

\section*{Data availability}
The ERA5 reanalysis \citep{hersbach_era5_2020} data was downloaded from the \cite{Copernicus_C3S_ERA5} store using the CDS API. The training weights, as well as the code to run FourCastNet-v2 and Pangu-Weather were obtained using the Open Source Earth2MIP package \citep{earth2mip}. CMIP6 historical simulation data can be obtained from the ESGF nodes via the following application (\url{https://aims2.llnl.gov/search/cmip6/}).

\section*{Code availability}
To produce AI-based forecasts with the FourCastNet-v2 and Pangu-Weather models, we use the Earth2MIP package \citep{earth2mip}, which allows us to run these models using already trained weights. All the necessary code to reproduce the results of the paper, including code to download and run the model simulations is accessible in GitHub (\url{https://github.com/bernatj/AI-based-attribution}).
% is attached in a ZIP file and will be made available online uppon acceptance to comply with the double-anonymised peer review guidelines. 

\bmhead{Declarations}
The authors declare that they have no competing interests.

\bmhead{Supplementary information}

Additional figures (S1 to S12) can be found in the Supplemental Information (SI) that accompanies this manuscript.

%Authors reporting data from electrophoretic gels and blots should supply the full unprocessed scans for key as part of their Supplementary information. This may be requested by the editorial team/s if it is missing.
%Please refer to Journal-level guidance for any specific requirements.

\bmhead{Acknowledgements}
This research work was supported by the Ministry for the Ecological Transition and the Demographic Challenge (MITECO) and the European Commission – NextGenerationEU (Regulation EU 2020/2094) through the CSIC Interdisciplinary Thematic Platform Clima (PTI Clima). DB and RGH were also supported by the EU-funded H2020 project CLINT (Grant Agreement No. 101003876).

%\section*{Author contribution}

%Some journals require declarations to be submitted in a standardised format. Please check the Instructions for Authors of the journal to which you are submitting to see if you need to complete this section. If yes, your manuscript must contain the following sections under the heading `Declarations':

%\begin{itemize}
%\item Conflict of interest/Competing interests (check journal-specific guidelines for which heading to use)
%\item Ethics approval and consent to participate
%\item Consent for publication
%item Data availability 
%\end{itemize}

%\noindent
%If any of the sections are not relevant to your manuscript, please include the heading and write `Not applicable' for that section. 


\bibliography{references}% common bib file



\section*{Supplementary material}

# Supplementary Information for the paper "AI-driven weather forecasts enable anticipated attribution of extreme events to human-made climate change"

B. Jiménez-Esteve*, D. Barriopedro, J. E. Johnson, and R. Garcia-Herrera
*corresponding author: bernatji@ucm.es.

## 1 Introduction

This supplementary material provides additional analysis to support the main findings presented in the main manuscript. For any clarification on the methods, please refer to the main text. Figures S1 to S6 show additional analysis for the Iberian heatwave of 2018. They include the daily evolution of the temperature at 2 m (T2m) field (Figure S1), and associated skill scores (model bias in Figure S2, and ensemble spread in Figure S3), the lead time dependence of the T2m bias (Figure S4), daily maps of the bias in geopotential height at 500 hPa (Figure S5), and the daily evolution of the ACC signal (factual - counterfactual) in T2m (Figure S6). Figure S7 shows the forecasted ACC signal for several parameters (minimum sea level pressure, storm location, storm size, precipitable water) of hurricane Florence in 2018. Figure S8 depicts the forecast uncertainty associated with the extratropical Ciarán. Figures S9 to S11 reproduce Figures 2, 4, and 6 of the main text but using forecasts generated by the Pangu-Weather AI-based weather model. Finally, Figure S12 shows the global attribution of T2m for a specific date with forecast initialisations from ERA5 and IFS analysis.

## 2 Supplementary Figures (S1 to S12)

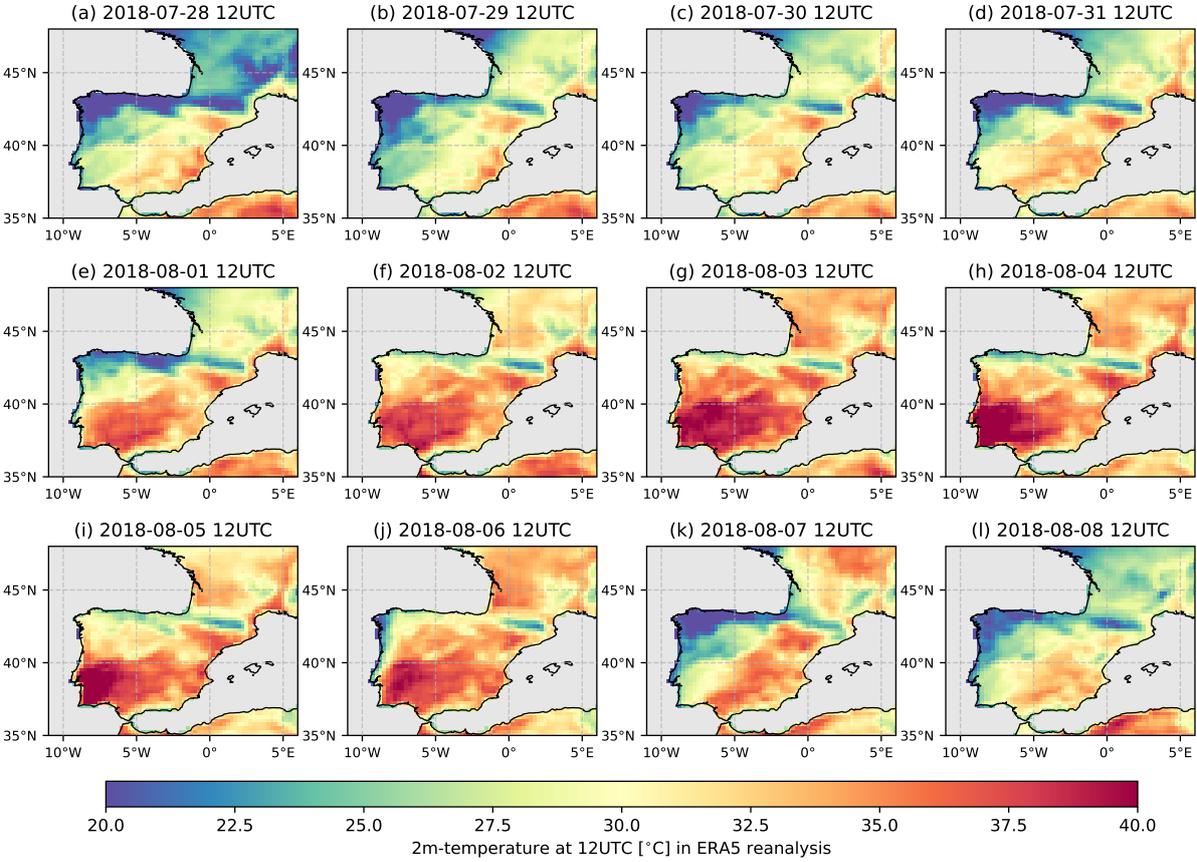

Figure S1: The 2018 Iberian heatwave. Daily evolution of 12UTC temperature at 2 m (T2m) during the Iberian heatwave of 2018 (°C). Data from ERA5 reanalysis.

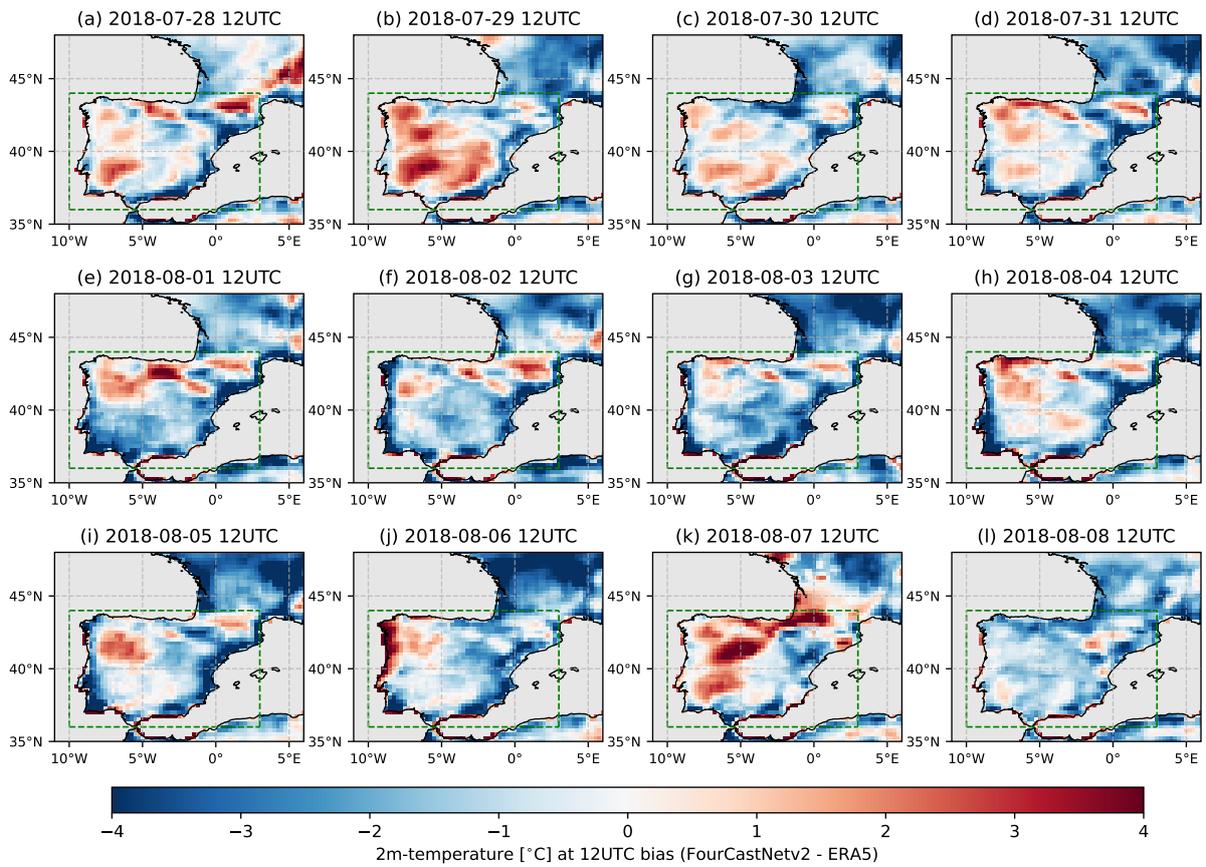

Figure S2: Forecast error of the 2018 Iberian heatwave. Daily evolution of the ensemble mean bias (FourCastNetv2 minus ERA5 difference) in 12UTC T2m (°C) during the Iberian heatwave of 2018.

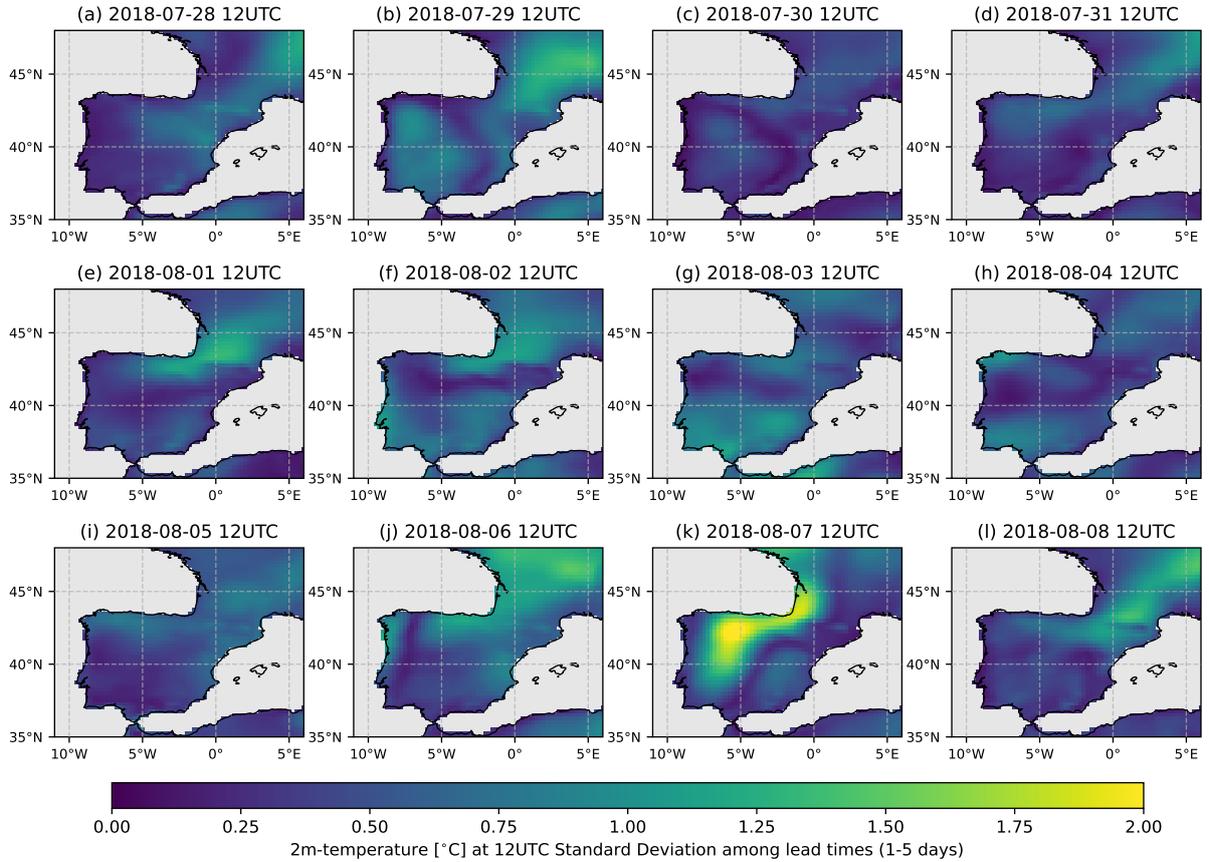

Figure S3: Forecast uncertainty of the 2018 Iberian heatwave. Daily evolution of the standard deviation of the 12UTC T2m (°C) predicted by the members of the ensemble during the Iberian heatwave of 2018.

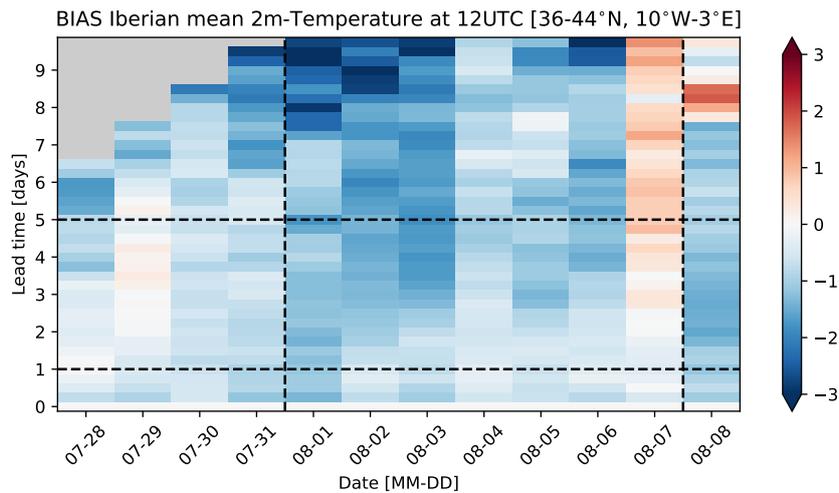

Figure S4: Lead time dependence of the bias for the 2018 Iberian heatwave. Area averaged [36-44°N, 10°W-3°E] bias in 12UTC T2m (°C) as a function of the date (x-axis) and lead time (y-axis). Vertical dotted lines indicate the start and end of the heatwave episode. The horizontal lines define the lead time interval employed to calculate the ensemble mean.

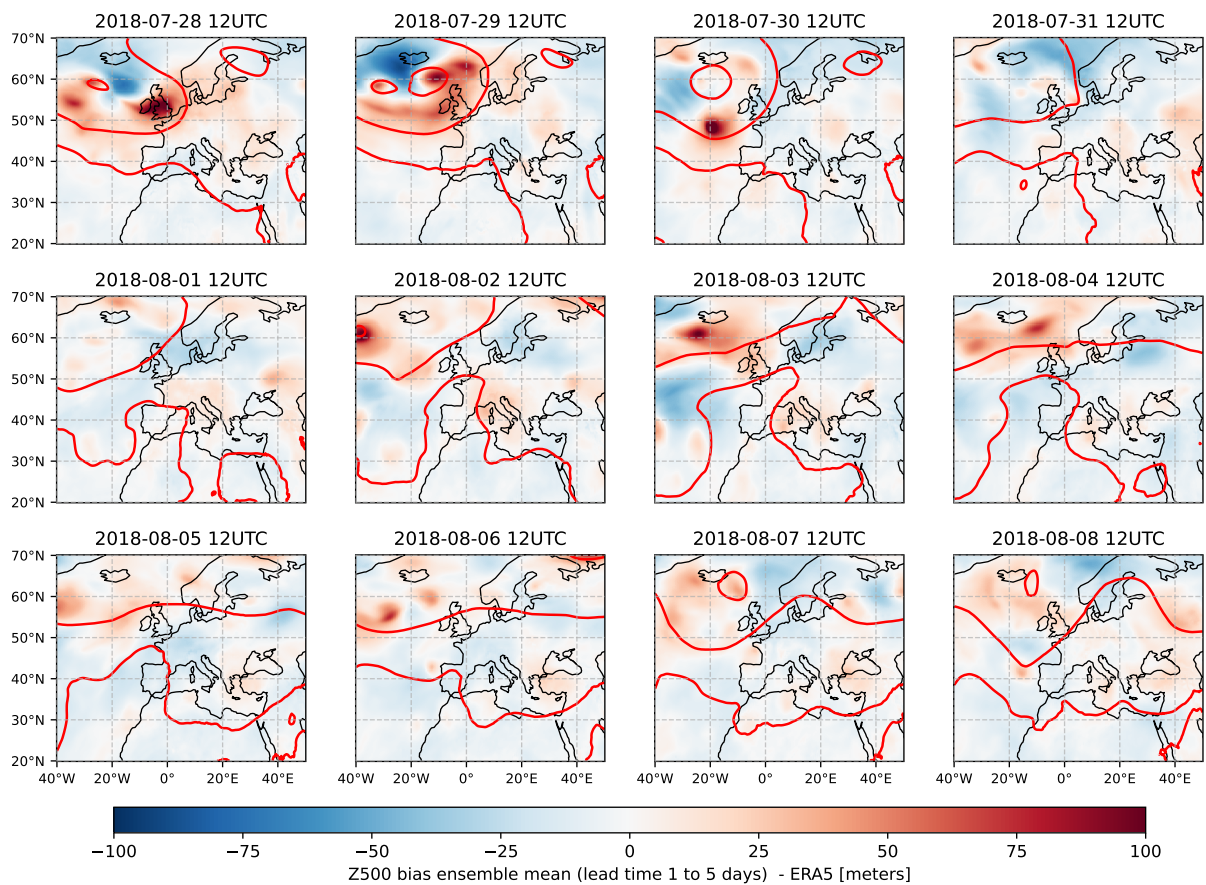

Figure S5: Atmospheric circulation during the 2018 Iberian heatwave. Daily evolution of the ensemble mean bias of geopotential height at 500 hPa (color shading, m). Red contours show the ensemble mean field (5400, 5700, 5900 m contours), defined as the average for lead times from 1 to 5 days.

5

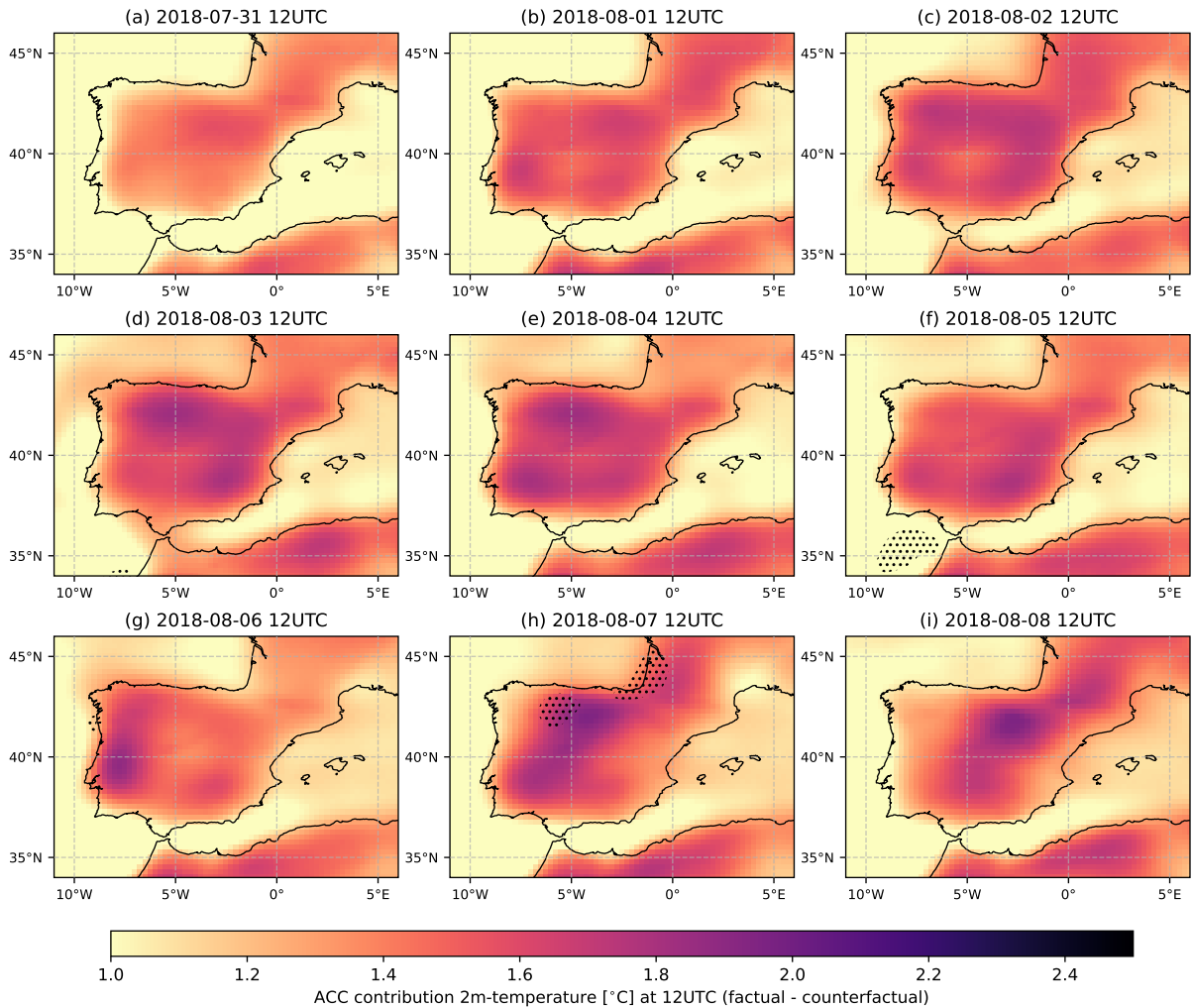

Figure S6: ACC attribution of the 2018 Iberian heatwave. Daily evolution of the ensemble mean ACC signal (factual minus counterfactual difference) in 12UTC T2m (°C). Non-significant values at the 95% confidence interval according to a two-sided t-test are dotted.

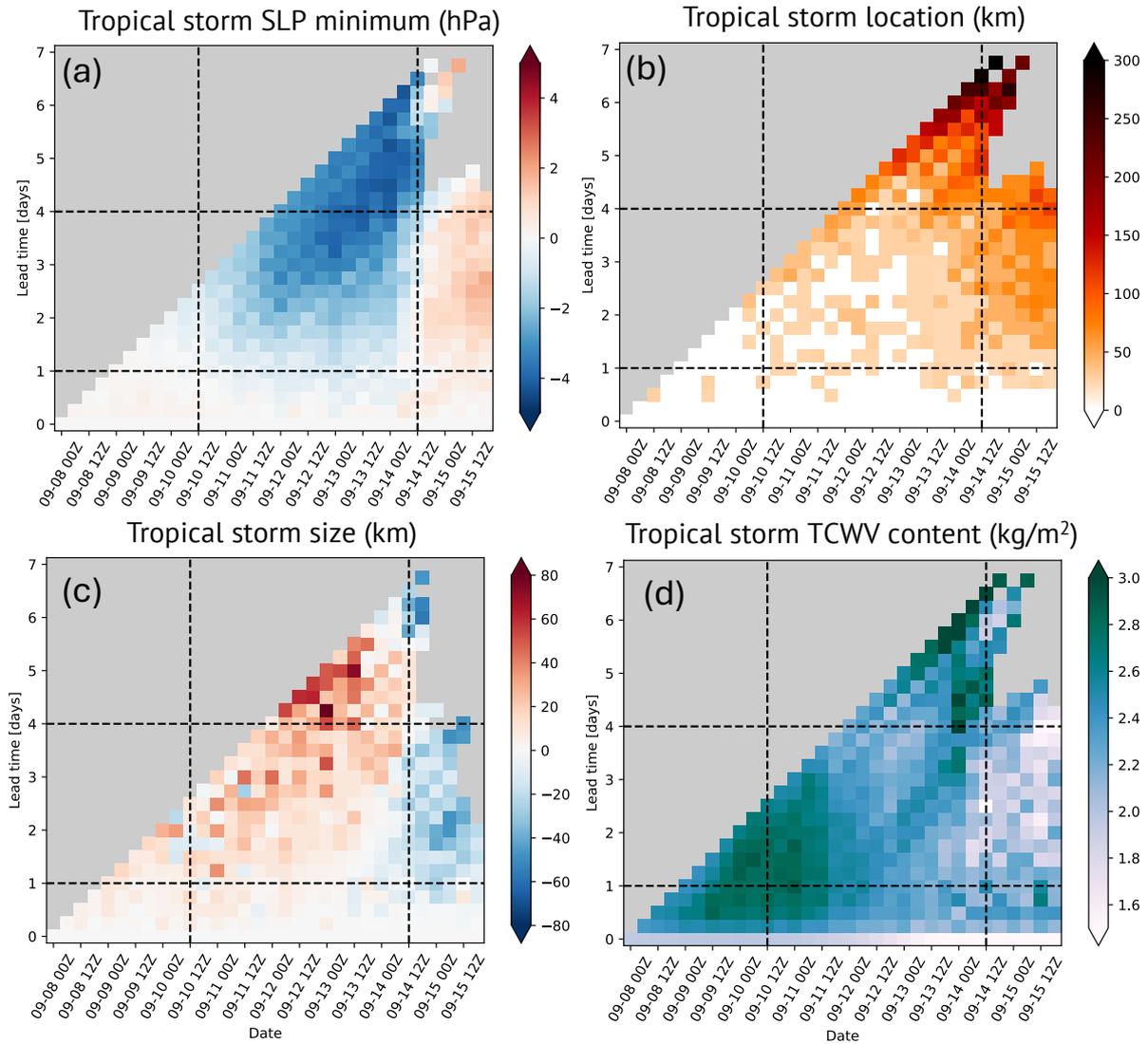

Figure S7: ACC attribution of hurricane Florence. ACC signals for hurricane Florence as a function of lead time (y-axis) and forecasted time (x-axis): (a) minimum sea level pressure (SLP) at the centre of the storm (hPa), (b) storm location (km), (c) storm size (km), and (d) precipitable water (kg/m$^2$) within 500 km from the tropical storm centre. Vertical dashed lines delimit the period between the largest intensification of the storm (12UTC 9 September 2018) and the landfall (12UTC 14 September 2018). Horizontal dashed lines define the lead time interval employed to create the lagged ensemble mean. See methods for definitions

7

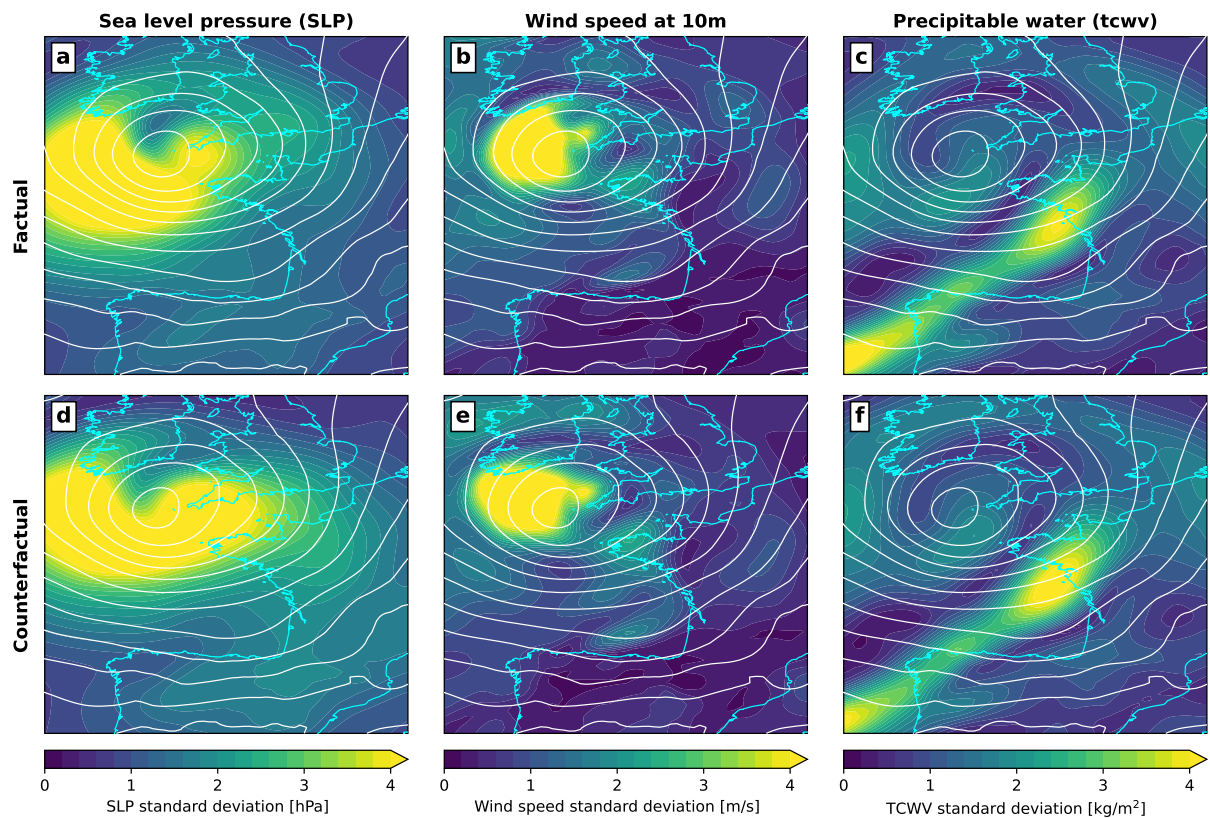

Figure S8: Forecast uncertainty of storm Ciarán. Ensemble spread calculated as the standard deviation over 1-to-4-day lead times (12 members) for the peak of the extratropical storm Ciarán (00UTC 2 November 2023) in the (a-c) factual and (d-f) counterfactual forecasts: (a,d) SLP (hPa), (b,e) wind speed at 10 meters (m/s), and (c,f) precipitable water (kg/m$^2$). White contour lines indicate the SLP field (hPa) forecasted by the ensemble mean.

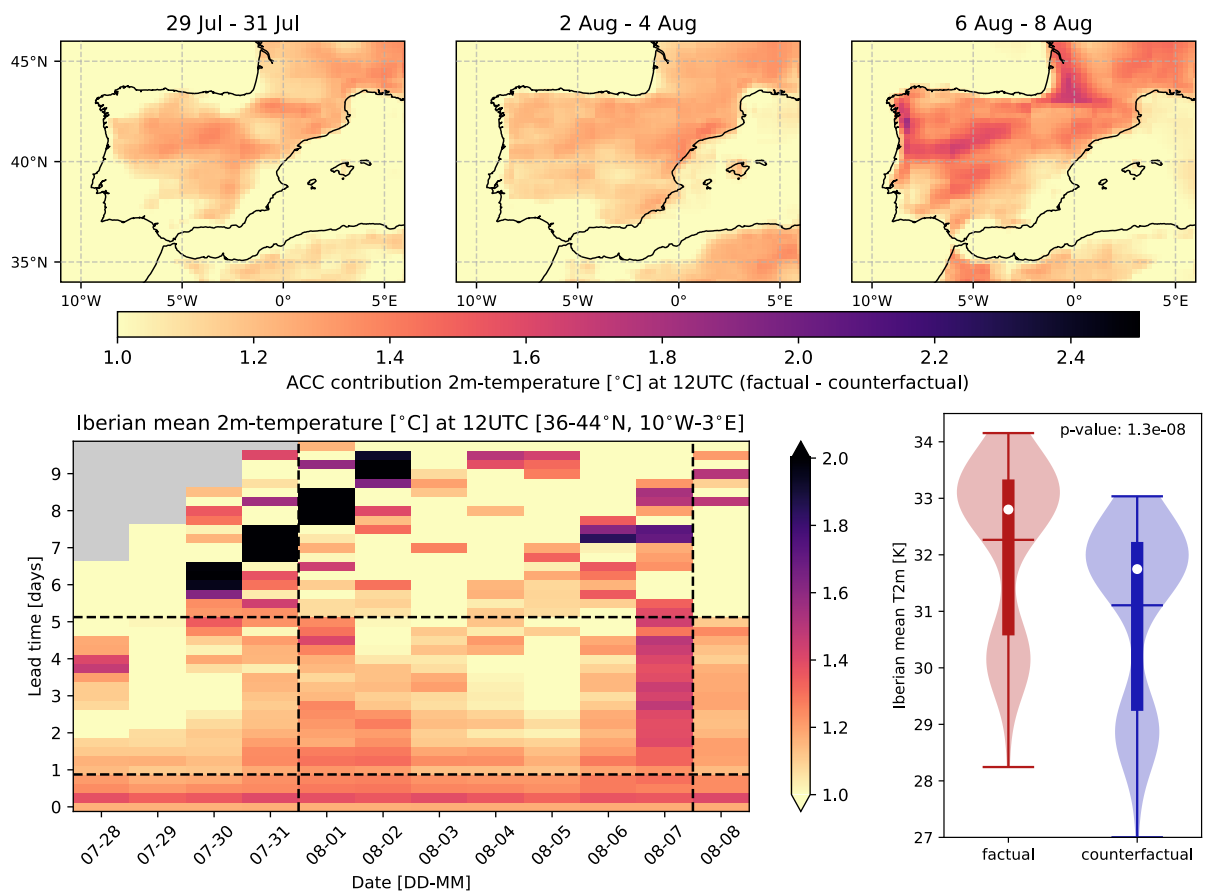

Figure S9: ACC attribution of the 2018 August Iberian heatwave. Same as in Figure 2 of the main text but using Pangu-Weather.

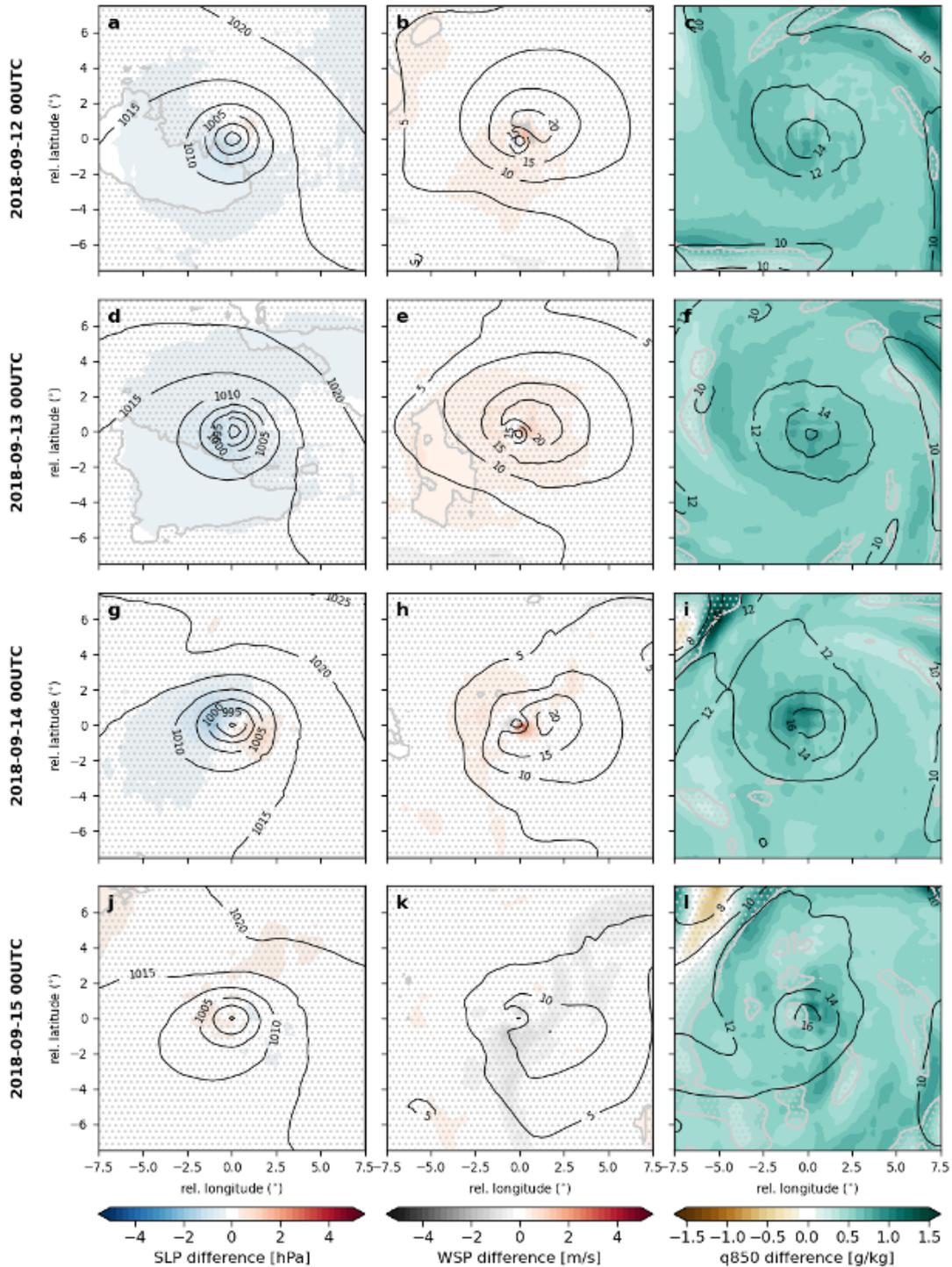

Figure S10: ACC attribution of hurricane Florence. Same as in Figure 4 of the main text but using Pangu-Weather. In (c,f,i,l) specific humidity (g/Kg) at 850 hPa is plotted instead of TCWV, as Pangu-Weather does not provide TCWV.

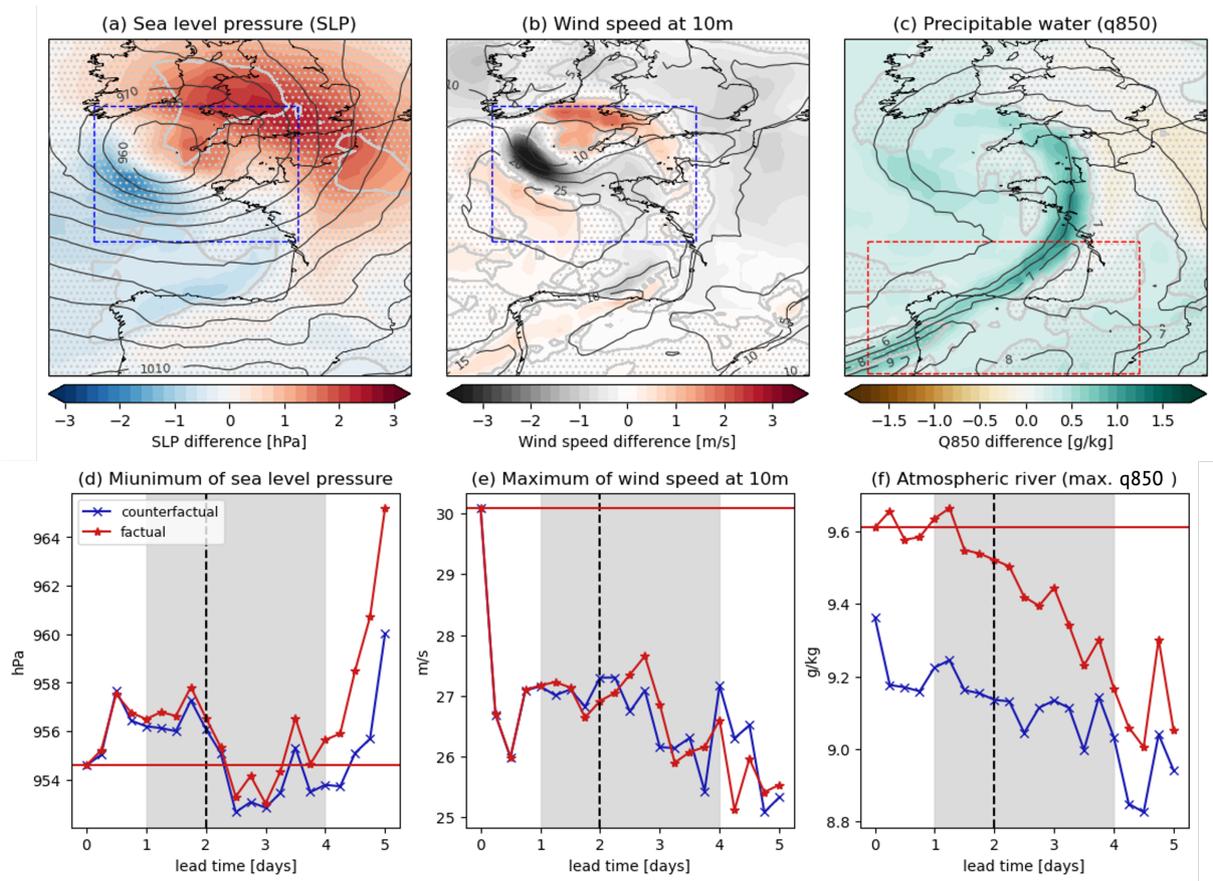

Figure S11: ACC attribution of extratropical storm Ciarán. Same as in Figure 6 of the main text but using Pangu-Weather. In (c) and (f) specific humidity (g/Kg) at 850 hPa is plotted instead of TCWV, as Pangu-Weather does not provide TCWV.

11



\end{document}